# Growing the Simulation Ecosystem:

## *Introducing Mesa Data to Provide Transparent, Accessible and Extensible Data Pipelines for Simulation Development*

Thomas D. Pike[a], Samantha Golden[a], Daniel Lowdermilk[a], Brandon Luong[b], Benjamin Rosado[b]

[a]NIU – C(AI)² Maryland, U.S., [b]Rutgers University New Jersey, U.S.

**Abstract:** The Agent Based Model community has a rich and diverse ecosystem of libraries, platforms, and applications to help modelers develop rigorous simulations. Despite this robust and diverse ecosystem, the complexity of life from microbial communities to the global ecosystem still presents substantial challenges in making reusable code that can optimize the ability of the knowledge-sharing and reproducibility. This research seeks to provide new tools to mitigate some of these challenges by offering a vision of a more holistic ecosystem that takes researchers and practitioners from the data collection through validation, with *transparent*, *accessible*, and *extensible* subcomponents. This proposed approach is demonstrated through two data pipelines (crop yield and synthetic population) that take users from data download through the cleaning and processing until users of have data that can be integrated into an ABM. These pipelines are built to be *transparent*: by walking users step by step through the process, *accessible*: by being skill scalable so users can leverage them without code or with code, and *extensible* by being freely available on the coding sharing repository GitHub to facilitate community development. Reusing code that simulates complex phenomena is a significant challenge but one that must be consistently addressed to help the community move forward. This research seeks to aid that progress by offering potential new tools extended from the already robust ecosystem to help the community collaborate more effectively internally and across disciplines.

## Watchmakers and Ecosystems

In his seminal work *Sciences of the Artificial*, Nobel Laureate Herbert Simon tells the parable of two watchmakers. Each watchmaker builds the same watch with 1,000 parts, the first breaks the watch into subcomponents of 10 pieces each, while the second painstakingly puts all 1,000 pieces together in an unbroken sequence. The first watchmaker prospers, while the second goes bankrupt. The reason for this disparity is every time the second watchmaker is interrupted, the watchmaker must start all over at part one of 1,000, while the first one only has to restart the 10 piece subcomponent (1). The parable is representative of a foundational aspect of computing, coding, and the coding ecosystems; the ability to have reusable code. The power of reusable code cannot be understated, but there is still a challenge with this parable. A 1,000-part watch is a complicated machine not a complex system. In a complex system, where the *whole becomes more than the sum of its parts*, you have the challenge of separating these components without losing the novelty of the system. In a watch you can take out a piece or subcomponent and put it back in and have a functioning watch. Conversely, for a living entity you cannot take out its heart, put it back in and get back a living entity. This dynamic epitomizes the challenge of having reusable subcomponents for simulations, specifically agent-based models, whose internal interdependencies create emergent phenomenon and are similarly difficult to decompose. There is extensive work across the ABM discipline, both explicit and implicit, that seeks to overcome or mitigate this fundamental obstacle. Building from this work, this research seeks to add to these existing ecosystems by offering a vision of a more holistic ecosystem that takes researchers and practitioners from the data collection through validation, with transparent, accessible, and extensible subcomponents. Ideally, this will aid the ABM community in more effectively storing, sharing, and supporting each other as our collective intelligence works to understand the vexing challenges of complex systems.

Prior to exploring the agent-based model ecosystem, however, it is necessary to start with a critical dynamic that drives modern computing applications, reusable code. Reusable code, and the related coding libraries, are ubiquitous in coding and they are subcomponents on which programmers and researchers depend. Like the watchmakers, instead of having to build each application from scratch, researchers and programmers leverage these reusable building blocks. The building blocks are stored knowledge that can be shared and improved. Besides just ease of assembly, subcomponents offer another critical advantage. Each library of code can be tinkered with independently, this maximizes the explore – exploit paradigm where attempted improvements can be made with minimal damage to the working whole (1,2). This is evident in the constant improvement of libraries associated with any programming language. As this dynamic is critical to programming, there is a robust and well-developed infrastructure to share and improve code. Coding repositories such as GitHub, Bitbucket, GitLab epitomize the ease at which multiple individuals can exploit the same piece of code and explore improvements. The benefit of subcomponents being a critical part of the programming is that there is well-developed infrastructure to host and develop simulation building blocks, so modelers can use, share, and reuse others code.

As the infrastructure exists and subcomponents are a fundamental part of the computing culture the remaining step from a modeler standpoint is to develop and store simulation subcomponents. However, the question of "What are the appropriate subcomponents?" is non-trivial, each simulation is a complex entity, and it is not obvious which parts should be subcomponents and how they should be stored and developed (3,4). To give a glimpse of the difficulty of this challenge, ABM modelers not only need reusable code but need stable subsystems. Critically, these subsystems are not static pieces, but have to be grown or evolved into a stable state to then interact with other stable sub systems in a model (5). The specifics of sharing and integrating such subsystems remains an open question and an area in need of research. An intended effect of this research is to enable such efforts. From this basic understanding, this research proceeds in three parts. First, a review of the different approaches to making the ABM ecosystem more accessible and extensible. This review will show where this research seeks to add to the literature, with the goal of taking the exceptional work that exists and furthering it. Second, this paper will discuss its effort to address the challenges of a dynamic ecosystem with the introduction of Mesa Data,[1] which provides two skill-scalable data pipelines (a crop yield pipeline and a synthetic population pipeline) as an attempt to facilitate more effective interdependencies. Critical to this effort is the ability for these data subcomponents to be compatible with existing and developed frameworks, so as the name implies these data pipelines are nominally connected to the Mesa[2] library (a python-based ABM library) (6)[3] and in the coding language Python. However, significant effort was taken to make the data itself accessible to non-Mesa/non-Python users as well as non or novice coders (hence the skill scalable designation). These efforts were taken to ensure maximize use, transparency, accessibility, and extensibility to each data processing decision in the data pipeline.[4] Third, the paper will discuss existing and future work to help further develop the ecosystem.

## The ABM Ecosystem

The ABM ecosystem is composed of a mix of different components from model repositories, to coding libraries to full applications. ComSES Net (Network for Computational Modeling in Social and Ecological Sciences)[5], which maintains a repository of models, lists 34 active modelling frameworks (3). The frameworks can then be generally broken down into four types. 1 - Libraries which allow users to code their own models providing ABM management modules such as schedulers and data collection. Some notable examples of these include MASON (7) , RePast (8), Agents.jl (9)  and Mesa (6). 2- Customized libraries which have been optimized for a specific purpose. Some examples of these include a vegetation-atmospheric ecosystem library SEIB-DGVM (10), a forest landscape model, LANDIS-II (11) and Epidemiological MODel EMOD through the Institute for Disease Modelling (12). There are also tangentially related simulations that are not necessarily ABMs, but could provide critical dynamics to aid ABM simulations, or more generically allow for the integration of different types of simulations. A review of such an approach is beyond the scope of this paper but as a related example, (13) examines water supply and demand for urban areas which could provide either data generation or validate simulation results.   3 - Platforms, which provide users great customization of their models by providing a unique high-level coding language and/or ready-made building blocks. Some notable examples include Evoplex (14) , GAMA (15), and Netlogo (16). 4- Applications, which require no coding, and can have software development kits and can be proprietary. Examples of this include AnyLogic (17) or Simudyne (18). In addition, to these frameworks are modelling repositories with ComSES maintaining a large repository of over 780 models and actively monitoring publications using ABMs with over 7500 publications (3). As well as NetLogo has a library of verified models,[6] user community models[7] and a modelling commons to share models[8], each representing repositories of stand-alone models (16). The ABM ecosystem has a rich diversity of approaches to storing the knowledge associated with building and implementing ABMs.

---

[1] https://github.com/projectmesadata
[2] https://github.com/projectmesa/mesa, https://mesa.readthedocs.io/en/master/overview.html
[3] To ensure full disclosure, the corresponding author Thomas Pike is a member of the Mesa Development team. However, in accordance with the Mesa operating procedures this nascent effort has not gained enough crowd – source support to ensure it sustainability to be considered an official part of Project Mesa. Volunteer contributors and maintainers are welcome.
[4] Users can explore the data pipeline by clicking on the BinderHub icon in ReadMe docs of the Mesa Data repository and opening one of the non-data download .ipynb files.
[5] https://www.comses.net/
[6] https://ccl.northwestern.edu/netlogo/models/index.cgi
[7] http://ccl.northwestern.edu/netlogo/models/community/index.cgi

The rich taxonomy of approaches represents the varied use of simulations that is consistent with the behavior of natural ecosystems where different species develop and symbiotically support each other and co-evolve. This research seeks to further develop the ecosystem by linking the generic libraries (i.e. MASON, Mesa), the customized libraries (i.e. LANDIS-II, EMOD) and the Platforms (i.e. Evoplex, GAMA) together. The initial step that intuitively seems the most promising is data pipelines. Data pipelines could ideally feed any simulation approach by outputting data that could be inputted or providing data for validation purposes. This has the added benefit of potentially being useful to other modeling approaches such as statistics or machine learning.

One challenge is that entities within complex systems are critically linked to other parts of the ecosystem (e.g. they are nearly decomposable). Due to this dynamic, this effort could not stand in isolation so had to be linked to existing parts of the ecosystem. In this case, the data pipeline uses Python and as noted earlier is linked to the a Python based Agent Based Model framework Mesa (6). However, based on some enabling coding technology the authors were able to aid accessibility with two design decisions. First, the code is designed to be skill scalable, in that users can traverse the data pipeline without code as well as see the code for each step. This was done was by using Jupyter Notebooks (19) with additional code and input widgets (20) so users can either ignore the code completely or explore the code (Figure 1). Furthermore, the repositories use Binder (21) or Google Colab (22), which will allow them to click the badge and see and explore the pipelines without dependency or software installation. Second, the pipelines walk step by step through the data conversion process so users can see the choices made and algorithms used. As information is relative not absolute, every data processing step comes with an information cost or decision that must be transparent to the user. The goal with these decisions is to maximize transparency and accessibility. Finally, with their presence on GitHub, individuals across the world can further improve and develop each data pipeline. With this understanding in mind, we encourage the reader to explore the following tools and provides us feedback or contributions through GitHub to improve the data pipelines or add others. The goal is these repositories live on their own and are sustained by the community.

---

[8] http://modelingcommons.org/account/login

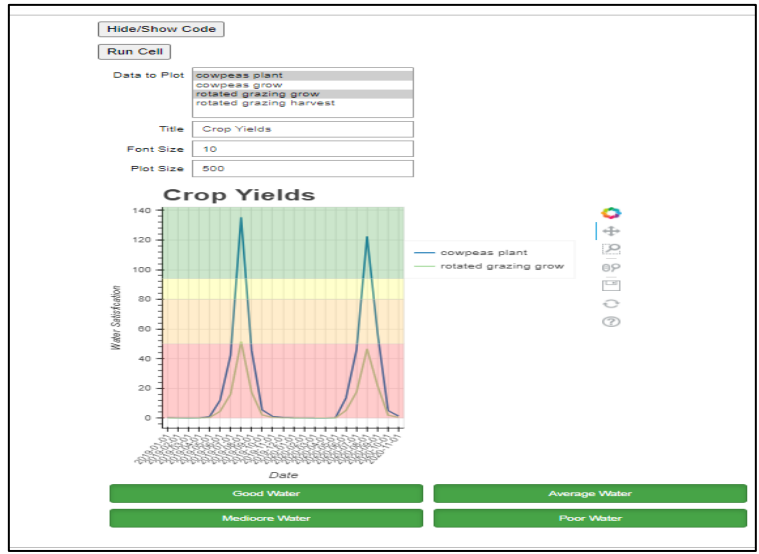

Figure 1: Example of interface on the maximal crop yield data pipeline. Based on previously selected crops, users can see a plot of maximal crop yield and select some or all the crops, see or hide the code, and adjust various parameters of the plot such as size or title.

## Mesa Data – Synthetic Population

This first pipeline is syntheticpopulation, designed to provide users a geolocated, demographically accurate synthetic population for their ABM. The population tool is split into four files (1) Population Data Download, (2) Density Exploration and Conversion, (3) Demographic Exploration and Conversion and (4) Synthetic Population Starter. The first three files are notebooks that proceed sequentially, describing each step, the data choices made, and the code performed. The population tool uses the WorldPop population and demographic datasets (23).  The synthetic population pipeline is available at on the Project MesaData syntheticpopulation GitHub repo[9]  or via the Binder Icon on the README page. The population tool provides users a data pipeline to create a geolocated, demographically accurate synthetic population for their ABM.

### WorldPop Datasets

The WorldPop dataset consists of 44,683 datasets as of  January 2021, which are used to provide detailed population outputs from a basic count to migration flows to urban change (23). The population tool uses two WorldPop outputs, the population count dataset and the demographic dataset to develop the synthetic population. The population count dataset is developed using remote sensing data, integrated with a Random Forest estimation technique that integrates census data to distribute the population at approximately 100-meter spatial resolution (24). Determining fine-grained demographic data provides additional challenges and requires a significant mix of techniques that generically mixes remote sensing data with census and survey data to gain insights into the large variance of demographics at different locations (25,26). WorldPop provides easily accessible, rigorous population and demographic data that is the foundation of the syntheticpopulation pipeline.

---

[9] https://github.com/projectmesadata/syntheticpopulation

## Creating a Synthetic Population

The population pipeline allows user to download population density and demographic data by country and then turns that data into tables of location, age groups and gender. The Density Exploration and Conversion file allows use to convert the downloaded density file from WorldPop into geolocated integers. Due to the nature of WorldPop Random Forest calculations its population density data output produces rational numbers spread across the country which is not conducive to creating a population of discrete agent objects. In addition, as the world population provides data by country, the density exploration and conversion file also provide users the ability to select a subset area of the country for their synthetic population. For example, a user can select just the capital of Tirana to build their synthetic population and not the whole country of Albania.  To address the issue of rational instead of integer numbers the tool sums the decimal portions that would be lost from rounding and then redistributes those whole numbers back into the largest numbers that would have been rounded down, as people are distributed via a pareto distribution (e.g. areas with lots of people get more people) (27,28). The effectiveness of this calculation varies based on the degree of accuracy selected by the user and the actual distribution of the population. World Pop provides nominal resolution to six degrees which is approximately equal to 0.11 square meters at the equator. Users can select from two degrees (~1.1-kilometer resolution) to six degrees. The lower the degree of precision the closer this method gets to the target population. In addition, errors magnify based on the size of the area being considered. In larger countries with significant areas of sparsely populated regions such as the sahel region of Africa, this approach can result in sparsely populated areas receiving no people instead of a few people. This result can be mitigated by selecting smaller regions to get the population, but the pipeline does not have a generalizable solution to turning the decimals in whole people. This reality of this weakness in the data pipeline highlights the main point of this effort. By placing this pipeline in a transparent, community sourced repository other individuals with specific knowledge or insights can improve this pipeline or add more transformation choices to share knowledge more effectively and provide users different options based on their unique concerns. Due to the size of these files the output from the density and exploration conversion file is saved in a hierarchical data format 5 (HDF5) file with latitude, longitude (to the desired accuracy) and integer population number. For ease of visualization, the web Mercator coordinates are also provided.  The Density Exploration and Conversion file converts the WorldPop population density data into a table of latitude, longitude and a discrete number representing the population.

Part three of creating a synthetic population is the Demographic Exploration and Conversion file. This file retrieves the demographic data consisting of the ages and gender at a specific location. The Demographic Exploration and Conversion file follows the same process as the Density Exploration and Conversion file and applies this to each age group provided by WorldPop (18 files of 9 different age groups for male and female).  After calculating the whole population, the demographic file uses the area selected in the density file to get the desired area from each of the demographic files. The demographic pipeline then reduces the area and uses the same process as used with the density file to get the integers for each population at a given area. A challenge is that if all the demographic population files are downloaded after conversion may be several gigabytes or more of data. For example, Niger which is approximately 1.2 million square kilometers is approximately 70 GBs of data. Using the HDF5 file format prevents memory errors but users may still need significant storage space. The Demographic Exploration and Conversion file takes each gender and age group file provide by WorldPop and converts them into integers with a specific latitude and longitude.

Part four the synthetic population starter python file provides an example of converting the demographic data into a synthetic population. The process is fairly straightforward, the code iterates over each table and then creates a Mesa Agent object assigning each agent a gender, age, latitude and longitude based on the file. The synthetic population starter provides users with some example code to convert the demographic data in agent objects with the associated attributes of the demographic files.

## Outputs and Visualizations

The population tool provides visualizations throughout to aid users in validating the data pipeline process and understanding what data is being produced. As the final output is a table that can easily be used to build agent objects, tables are displayed throughout the pipeline process. However, additional visualizations help portray the data. For the Density Exploration and Conversion, a heat map is produced of data to show where in the country are the key population centers (Figure 2). This map is also used to help users select a specific area of the country instead of having to retrieve the population of the whole country. The output from the Density file is a HDF5 file continuing the latitude, longitude, web Mercator latitude, web Mercator longitude and each location respective population.

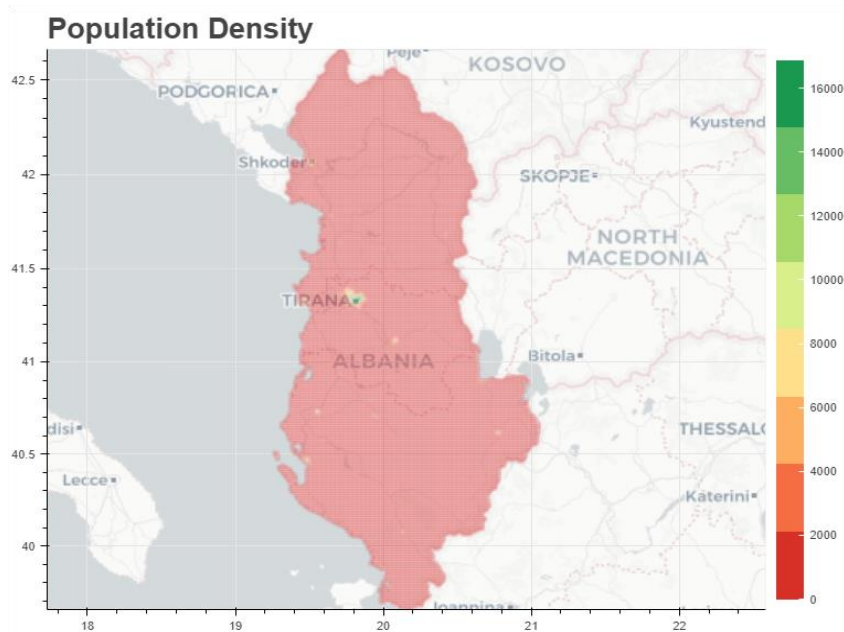

Figure 2: A heat map of the population density of Albania. This a visualization from the synthetic population data pipeline. Its purpose is to allow users to visualize and better understand the data. It also provides the option to down select to a specific area of interest.

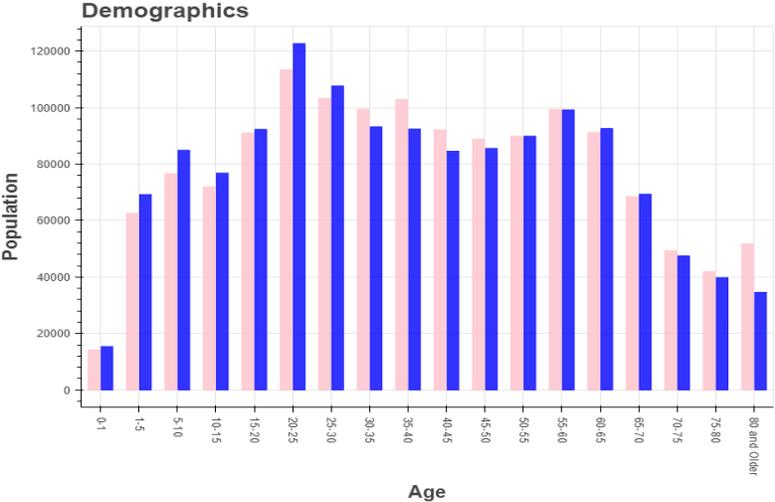

Figure 3: A demographic bar chart of Albania 2020. This is another visualization from the synthetic population data pipeline that helps users explore and understand the data process.

The Demographic Exploration and Conversion file produces a vertical bar graph of the age groups and genders so users can see their population make up for the desired year. (Figure 2) The demographic file also produces a heat map of the population so users can see the heatmap of a specific area. This provides a visual validation of the data to ensure there is not an error somewhere in the download and transformation process or missing or corrupted data. The output for Demographic data pipeline is a HDf5 files with a group for each demographic with the latitude, longitude, and integer population. This can then be easily iterated through to develop a realistic synthetic population for a simulation.

## Mesa Data – Crop Yield

The second pipeline is cropyield, specifically this pipeline calculates the maximal crop yield for selected crops based on the water satisfaction requirement index (WSRI). The purpose is to have a modular data pipeline that feeds in the environment of the ABM, while also allowing users to explore and understand the data. The crop yield data pipeline is split into three Jupyter notebook files: (1) Crop Yield – Data Download, (2) Crop Yield Location and (3) Crop Yield Regional. Each file proceeds sequentially, describing each step and the actions taken with hyperlinks to relevant references, keeping with the goal of providing maximum transparency and modularity for user development and improvement. The tool uses three free open-source datasets and is processed using a crop forecasting algorithm which outputs an array of maximal yield crop growth which can be feed into an ABM and outputs the data in a table and various visualizations to optimize user understanding of the data (Figure 4). To ensure ease of exploration the data for Niger, a highly agricultural dependent society, is stored on the GitHub repo and a Binder instance created so interested parties can easily explore Crop Yield Location and Crop Yield Regional via their web browser. This instance available via the Binder Icon on README page. The crop yield tool provides an easy to use, transparent pipeline to integrate crop yield data into ABMs.

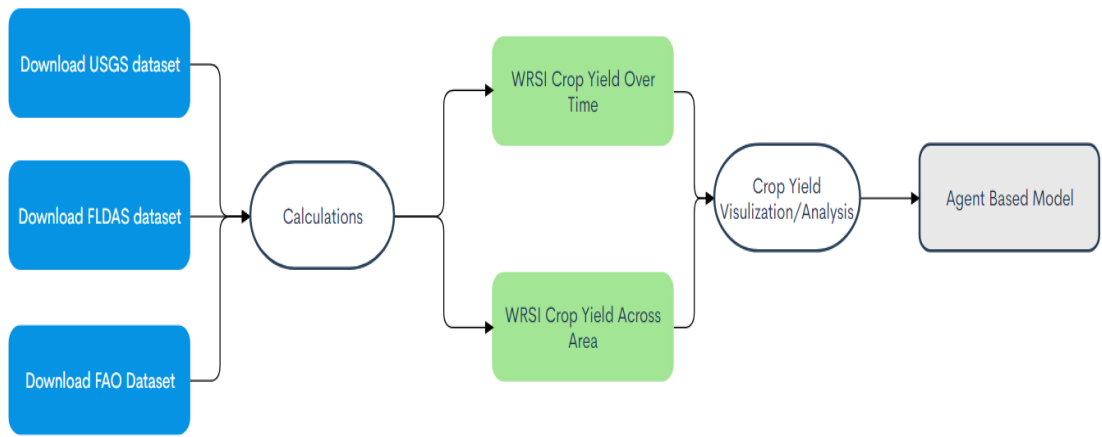

Figure 4: Flow Chart of Crop Yield data pipeline. Crop yield combines data from two NASA data sources, one United Nations reference. Cropyield provides users the option of seeing data at a specific location over time (to provide users higher fidelity of what the data provides) or seeing the data across a region for a specific time.

## The Datasets

Two datasets and a reference document are necessary to assess crop yields. The datasets are from NASA's Earth Data repository and although free, require a user account. The NASA datasets are (1) The Famine Land Data Assimilation System (29) and (2) the United States Geological Survey (USGS) Elevation data (30). The reference document is *Crop Evapotranspiration – Guidelines for Computing Crop Water Requirements* from the United Nations and provides water requirements for various crops at different stages of growth (31).

Although each data source is produced through highly technical processes that goes beyond the scope of this paper, a brief overview is necessary to understand the information they provide and how it is processed. FLDAS is a special instance NASA's Land Information Systems (LIS) and provides 30+ years of monthly data and a wide range of information to assess famine conditions (29). The crop yield tools does not use all the data FLDAS provides, instead it retrieves the air temperature, humidity, net short radiation, net long radiation, wind speed and evapotranspiration. These data dimensions are required to conduct the crop yield calculations discussed in the next section. The elevation data is straightforward in that it retrieves the elevation of the inputted area to approximately 1 kilometer accuracy. Paradoxically, the download process for this dataset takes the longest, but produces a very compact .csv file of latitude, longitude and elevation. The final data source the *Crop Evapotranspiration* reference, provides the water requirements for various crops at three different parts of their life cycle (when they are planted, when they are grown, and when they are harvested). Although the tools currently provide options to calculate the water requirements for numerous crops that are critical in Niger, it is a fairly simple process to add more as necessary. These three data sources provide the necessary data to calculate the maximal crop yields of a given area.

## Calculating Maximal Crop Yield

The maximal crop yield is calculated using the Penman-Monteith algorithm (31) in the Python Crop Simulation Environment (32) to determine the Water Satisfaction Requirement Index (WSRI). WRSI calculates the land's fertility for a crop based on the water supply and demand that a crop needs to grow at different stages in its life cycle. It is calculated as the ratio of seasonal actual evapotranspiration (AET) to the seasonal crop water requirement (WR).

$$WRSI = \frac{AET}{WR} * 100 \tag{1}$$

Where AET is the actual measured seasonal evapotranspiration. WR is the crop water requirement and can be calculated using the equation:

$$WR = PET * K_c \tag{2}$$

Where PET is the potential evapotranspiration calculated using the Penman-Monteith potential evapotranspiration equation and $K_c$ is the crop coefficient, which changes based on the crop and the growth stage of the crop. $K_c$ is found in the United Nation's FAO's crop data source. FLDAS provides the necessary data, the air temperature, humidity, net short radiation, net long radiation, wind speed and evapotranspiration, to calculate the AET and PET. Calculating the PET further requires the elevation data obtained from the APPeears dataset. This function took the inputs of date, latitude, elevation, air temperature, net shortwave radiation, vapor pressure, and wind speed to calculate PET through the Python Crop Simulation Environment. The maximal crop yield is determined by calculating the Water Requirement Satisfaction Index (WRSI) using the Penman-Monteith algorithm calculated with the Python Crop Simulation Environment.

## Outputs and Visualizations

The Crop Yield Location and Crop Yield Regional files produce visualizations and outputs of the final calculations. The Crop Yield Location produces a time series of plot of the WSRI for the location and crops selected (Figure 4). The user can manipulate several parameters of the plot via interactive inputs. The primary purpose of Crop Yield Location is to show users what is happening at each location and how the values vary over time. This file is less useful for developing synthetic terrain in ABMs, but was developed to ensure greater user insight into the processes the data pipeline uses (*transparency* and *accessibility*). The Crop Yield Regional files produces a heat map of the area for a given month (Figure 5).

Users can alternate between the month they calculate to see how the WSRI changes over selected times. Of note, the Crop Yield Location walks users step by step through each process from data to output. While the Crop Yield regional performs those same calculations as every ~1.1 KM location to produce the heat map. This output is then placed in a .csv file for each month selected and can be used to populate the synthetic terrain of an ABM. The crop yield files, visualization and outputs are intended to maximize users understanding of the process while reducing the total time cost required to understand the process and initiate a synthetic terrain.

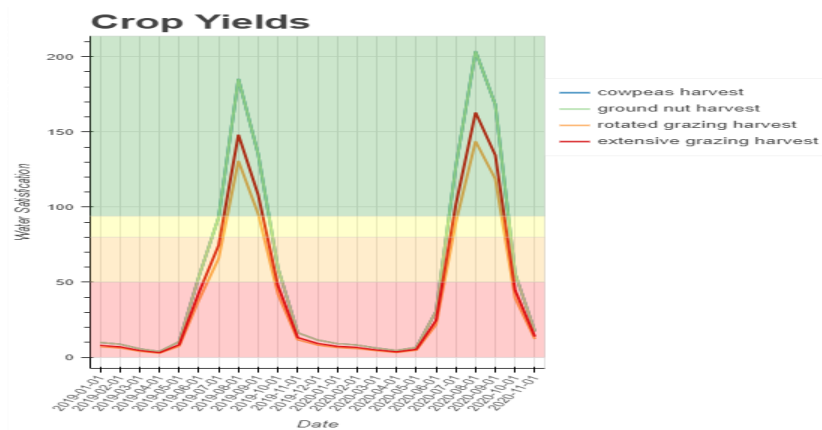

Figure 4: Time series plot of water satisifaction requirement index based on user selected crops and location.

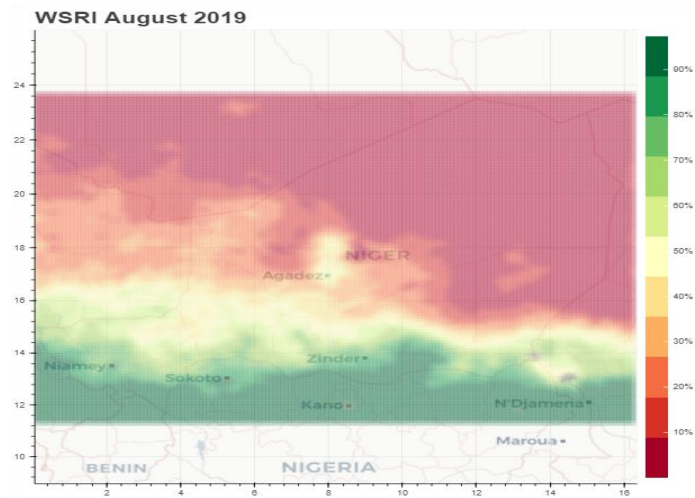

Figure 5: Regional heat map that provides an image of the water satisfaction requirement index for a one month period. Output of data for each month is saved in a table to aid the development of syntehtic terrain.

The crop yield tool tool provides a transparent and accesible data pipeline so users can see from data collection to output each step taken while requiring zero python but allowing users to see and change the code if desired. This tool uses well respected data sources and user inputs to transform the data into a format compatible for ABMs. The tool produces a time series graph for a specific location and a regional heatmap to maximize user understanding of the data and the processes used. Finally, the tool produces a series of .csv files that users can easily read into agent attributes as they initialize their ABMs. The cropyield repository is a comprehensive customizable data pipeline to support time series crop yield data ingestion into ABMs.

## Next Steps

The synthetic population and crop yield tool provide two examples of a way to further develop the robust ABM ecosystem to provide more transparency, accessibility, and extensibility to users. These pipelines are one small step and are only successful if they provide enough value added to the community that the community aids in their development and maintenance. Substantial challenges still remain in the optimal ways to integrate agent behaviors as well as the more challenging development of reusable stable subsystems (5). Further adding to this challenge will be the need to try and make these libraries more widely compatible so they are as language and operating system agnostic as possible. Despite these challenges efforts are underway to further develop more re-usable subcomponents. In Mesa Data, the authors have started a nascent pipeline to aid in validation of civil violence simulations using the Armed Conflict Location Event Data (33,34). In addition, a project to capture reusable agent behaviors has been started in Mesa Behaviors(35). These include an SIR (Susceptible, Infected, Recovered) (36) algorithm for epidemic modeling, the Bilateral Shapley Value which provides a coalition game theory algorithm as a behavior for group formation (37) and Multi-Level Mesa which facilitates networked agent populations which allows for active and dormant links. The great challenge with all this is that building such infrastructure is a laborious task. Thanks to the dynamics of crowd source coding and the connectivity provided by the internet, modelers can work across the globe to further develop the ABM ecosystem. The underlying question being "what are the key ingredients to start and incentivize this development, so it becomes a self-sustaining ecosystem?"

Mesa Data is an effort to find new dynamics to further aid the development of a robust simulation ecosystem for Agent-Based Models. Due to the inherent complexity of simulations this ecosystem is more difficult and complex than other thriving knowledge sharing ecosystems, such as machine learning libraries. This difficulty is based on the number of disparate parts that work together and the need for reusable stable subsystems (5) to provide a verified and valid simulation. This research seeks to aid the evolution of this ecosystem and help link more parts of the robust ABM ecosystem together. Ideally this will help further democratize simulation integration and use across researchers and practitioners helping the community to explore the hard challenges of complex systems more effectively.